\documentclass[10pt,oneside]{article}
\usepackage{graphicx}
\usepackage{amsmath}
\usepackage{amsfonts}
\usepackage{amssymb}
\usepackage{bm}
\usepackage{float}

\begin{document}

\begin{center}
\vspace{5cm}

\large\textbf{TERAHERTZ PHOTOGALVANIC SPECTROSCOPY \\
	OF  THREE DIMENSIONAL TOPOLOGICAL INSULATORS}

\vspace{1.5cm}
HELENE PLANK

\textit{\small{Faculty of the Institute of Experimental and Applied Physics, University of Regensburg \\
	Universitaetsstrasse 31, 93053 Regensburg, Germany \\
	helene.plank@ur.de}}

\vspace{0.5cm}
SERGEY D. GANICHEV

\textit{\small{Faculty of the Institute of Experimental and Applied Physics, University of Regensburg \\
	Universitaetsstrasse 31, 93053 Regensburg, Germany \\
	sergey.ganichev@ur.de}}

\vspace{1cm}
\end{center}

The paper overviews experimental and theoretical studies of photogalvanic effects induced in BiSbTe- based three dimensional topological insulators by polarized terahertz  radiation. We present the state-of-the-art of this subject, including most recent and well-established results.
We discuss a phenomenological theory based on symmetry arguments and models illustrating the photocurrents origin. 
We give a brief glimpse of the underlying microscopic theory, as well as an overview of the main experimental results. 

\vspace{0.5cm}

\vspace{0.5cm}
\textbf{1. Introduction}
\vspace{0.1cm}

Topological insulators (TI) have attracted great attention in the last years, for reviews see Refs.~[1,2]. 
A particular property of TI materials is, that they are characterized by an inverted bulk band structure and topologically protected  surface states. 
As a consequence, the two dimensional energy dispersion is described by the zero mass Dirac equation. Furthermore, for non-trivial TIs with an odd number of Dirac cones in the Brillouine zone (typically one) the system is characterized by  spin-momentum locking. The latter yields challenging fundamental concepts and makes TIs attractive for novel applications in the fields of spintronics and opto-electronics.
Currently, a variety of TI  systems has been theoretically predicted  and experimentally realized. 
Prominent examples for three dimensional (3D) TIs are BiSbTe- and BiSe- or strained HgTe-based thin films
and for two dimensional (2D) TIs are HgTe or GaSb/InAs quantum wells, for review see Ref.~[3]. 
To gain insight into the electronic properties of these materials, angle-resolved photo emission spectroscopy (ARPES)\cite{ARPES_1} 
and magneto-transport measurements\cite{Bardarson2013} are widely used. 

Recently a  further approach to study Dirac fermion systems has been developed and 
applies various phenomena induced by the high order of radiation electric fields\cite{GlazovGanichev_review,Ivchenko2017}.
One of the most often used methods to study TIs is based on photogalvanic effects (PGE), which scale as a square of the radiation electric field. Two types of photogalvanic effects have been observed and are extensively studied in TI materials: \textit{dc} photocurrent in response to high frequency radiation excited by linearly polarized radiation called linear PGE\cite{Olbrich2014}$^-$\cite{Plank2017}
and photocurrents sensitive to the radiation helicity being called circular PGE\cite{HosurBerry}$^-$\cite{Dantscher2017}.  
%
The variety of photogalvanic effects is completed by
magnetic field induced photocurrents gathered in the class of
magneto-photogalvanic effects\cite{Ivchenko2017,Eroms} and observed in graphene\cite{Drexler2013} and HgTe-based 3D TIs\cite{Dantscher2015}.

In this paper we give an overview on photogalvanic effects excited in BiSbTe- based 3D TIs. 
We focus on photocurrents excited by terahertz radiation and limit the discussion to the results on the linear PGE (LPGE), because in these materials and for this frequency range the circular PGE (CPGE) has not been detected so far. 
We demonstrate that in these materials  photogalvanic effects  can only be excited in the non-centrosymmetric surface states due to symmetry arguments. 
This allows the study of electron transport in surface states even in materials with a large number of residual impurities and at room temperature; conditions for which magneto-transport measurements are hardly applicable. 

We present a phenomenological analysis based on symmetry arguments, outline the microscopic theory based on the Boltzmann kinetic equation and give an overview of the main experimental results. 
We also address the role of another second order high frequency transport phenomena - the photon drag effect, which in contrast to the PGE, can be excited in both, surface and bulk states in the considered 3D TIs\cite{Plank2016}.
%
We show that in very different BiSbTe systems normal incident terahertz radiation results in a dominating contribution of the LPGE providing the basis for photogalvanics spectroscopy of Dirac fermions in the surface states. 
In particular, the LPGE spectral measurements in the terahertz range, in which the current formation is caused by  Drude-like absorption, allows one to determine room temperature carrier mobilities in the surface states\cite{Plank2016_2,Plank2017}.
Furthermore, measuring the polarization dependence of the photogalvanic current and scanning the beam spot across the sample, provides an access to topographical inhomogeneity's in the electronic properties of the surface states and the local domain orientation\cite{Plank2016_2}.

\vspace{0.5cm}
\textbf{2. Phenomenological description}
\vspace{0.1cm}

Photocurrents caused by homogeneous excitation of homogeneous materials, which is because of the weak radiation absorption usually the case in the terahertz range , are phenomenologically described by writing the   current  as an expansion in powers of
the electric field $ \bm E(\omega )$ at the frequency $\omega$ and the photon momentum inside the medium
$\bm q$\cite{Ivchenkobook2,book}. 
The lowest order nonvanishing terms yield a $dc$ current density $\bm j$, given by
\begin{eqnarray}
	\label{phenom} 
	j_{\lambda} =\sum_{\mu, \nu} \chi_{\lambda \mu \nu} E_{\mu} E^*_{\nu} 
	+ \sum_{\delta, \mu, \nu} {\cal T}_{\lambda \delta \mu \nu} q_{\delta} E_{\mu} E^*_{\nu} + c.c., 
\end{eqnarray}
in which  $\bm E$ is the electric field and  the expansion coefficients $\chi$ and ${\cal T}$
are third and fourth rank tensors, respectively.
In Eq.~(\ref{phenom}) the first sum represents photogalvanic effects whereas  the second sum describes photon drag effects containing additionally the photon momentum.

\vspace{0.5cm}
\textbf{3. Photocurrents excited at normal incidence}
\vspace{0.1cm}

\begin{figure}[t]
	\includegraphics[width=\linewidth]{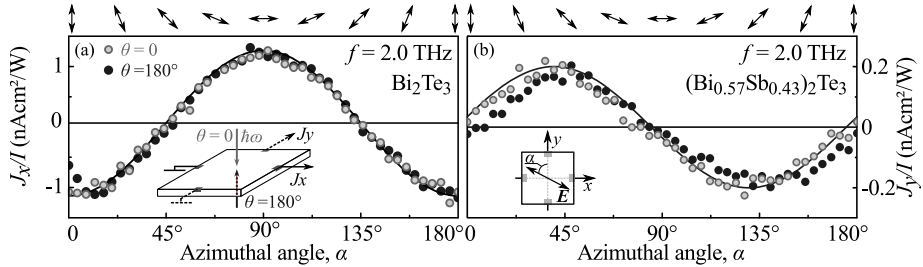}
	\caption{
		Photocurrent $J_{x}$  measured  in Bi$_2$Te$_3$, panel (a), and  $J_{y}$  in a (Bi$_{0.57}$Sb$_{0.43}$)$_2$Te$_3$ sample, panel (b), both excited at $f = 2.0$~THz and normalized on radiation intensity $I$. 
		Fit after Eq.~(\ref{phenom}). 
		Data are shown for both, front ($\theta = 0$) and back ($\theta = 180^\circ$) illumination. 
		Arrows on top schematically show orientation of linear polarization. 
		Insets show experimental setup and define angle $\alpha$.  	
		After Ref.~[9]. 
	}
	\label{figure1}
\end{figure}
%
Considering the  C$_{3 {\rm v}}$ point group symmetry of the surface in BiSbTe-based 3D TIs
and radiation at \textit{normal} incidence, Eq.~(\ref{phenom}) reduces to\cite{Olbrich2014} 
\begin{eqnarray}
\label{normal}
j_x &=&   (\chi + {\cal T} q_z)  [|E_{x}|^2-|E_{y}|^2] = - (\chi + {\cal T} q_z) E_0^2  s_{\rm 1}  \\
j_y &=&   - (\chi + {\cal T} q_z) [E_{x} E_{y}^*+E_{y}E_{x}^*] = (\chi + {\cal T} q_z) E_0^2  s_{\rm 2},  \nonumber 
\end{eqnarray}
with $x$- and $y$-axes  directioned along and perpendicular to one of the mirror reflection planes 
of the C$_{3 {\rm v}}$ point group and $s_{\rm 1} = \cos(2\alpha)$ and $s_{\rm 2} = \sin(2\alpha)$ being 
the Stokes parameters of light for linear polarization. Here the azimuthal angle $\alpha$
is the angle between the orientation of the radiation electric vector $\bm E$ and $y$-axis. 

A photocurrent excited by \textit{normal} incident linearly polarized radiation 
was detected in  very different BiSbTe- based 3D TIs in a wide frequency range from fractions of THz
up to tens of THz. It is characterized by the same overall behavior: 
in agreement with Eq.~(\ref{normal}) it scales quadratically with the radiation electric field, 
has a response time of picoseconds or less,  and exhibits a characteristic polarization dependence.
Figures~\ref{figure1} (a) and (b) show exemplary the polarization dependence of the photocurrent measured in $x$-direction for pure Bi$_2$Te$_3$ and in $y$-direction for (Bi$_{0.57}$Sb$_{0.43}$)$_2$Te$_3$, respectively.
As addressed above and as it clearly follows from  Eq.~(\ref{normal}), photogalvanic effects require the absence of an inversion center. Thus they can be excited in the surface states only, whereas photon drag effects are also allowed in the  bulk, which is centrosymmetric in BiSbTe-based 3D TIs. 
Olbrich et al. suggested a recipe allowing to distinguish between  photogalvanic and photon drag contributions\cite{Olbrich2014}. 
It is based on the fact that, besides the radiation electric field components, the photon drag is also proportional to the photon momentum. 
Therefore, for front ($\theta=0$) and back ($\theta=180^{\circ}$) illumination the LPGE current remains unchanged while the photon drag current reverses its sign. 
Here $\theta$
is the angle of incidence. Experiments performed on a large variety of materials reveal that in terahertz 
frequency range and \textit{normal} incident radiation 
the LPGE usually dominates the photoresponse. 
Typical results are shown in Fig.~\ref{figure1},  clearly demonstrating that photocurrents for front and back illumination are almost indistinguishable.
This proofs unambiguously that the photocurrent is caused by the LPGE and, consequently, is excited in the topologically protected surface states only. 

The generation process of the LPGE can be illustrated as follows: 
The applied terahertz radiation causes an \textit{alignment} of carrier momenta
along the direction of the radiation electric field.
%
The alignment corresponds to a \textit{stationary} correction to the electron distribution function, which scales as a square of the $ac$ electric field magnitude\cite{Olbrich2014}.
The alignment itself, however, does not lead to a $dc$ current. The mechanism of the current formation involves
asymmetric scattering of carriers by trigonal scatters. For the surface states in BiSbTe- based 3D TIs
the scatterers can be considered as randomly distributed but identically oriented wedges,  preferentially  along the crystallographic axes, see Ref.~[8,9]. 
Complementary $X$-ray measurements confirmed the three-fold symmetry of the surface
states. 
Due to this asymmetric scattering by wedges, the excess of the flux of Dirac fermions moving back and forth along the external $ac$ electric field results in a $\textit{dc}$ photogalvanic current, for details see Refs.~[8,9]. 

%
\begin{figure}[t]
	\includegraphics[width=\linewidth]{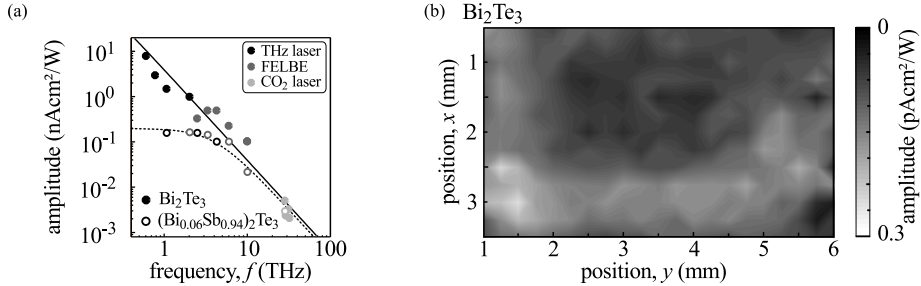}
	\caption{
		(a) Frequency dependence of LPGE amplitude measured in Bi$_2$Te$_3$ (full circle) and  (Bi$_{0.06}$Sb$_{0.94}$)$_2$Te$_3$ 
		(open circle). Fit after Eqs.~(\ref{chi}) and (\ref{Drude}).
		(b) LPGE amplitude as a function of coordinate for Bi$_2$Te$_3$. The data are obtained by scanning with the laser radiation of $f = 28$THz and spot of 30~$\mu$m in an 2x5$mm^2$ area. 
		After Refs.~[10] and~[11]. 
	}
	\label{figure2}
\end{figure}
%
For the terahertz frequencies discussed here the photon energies are much smaller than the Fermi energies. 
Therefore, a semi-classical approach can be applied for the calculation of the underlining above 
mechanism of the LPGE current generation.  The kinetic Boltzmann equation for 
the electron distribution function $ f_{\bm p}(t)$ is given by\cite{Olbrich2014}
%
\begin{eqnarray}
\label{BoltzmannPGE}
{\partial f_{\bm p} \over \partial t} + e \bm E \cdot {\partial f_{\bm p} \over \partial \bm p}
= - \sum_{\bm p'} \left( W_{\bm p', \bm p} f_{\bm p} - W_{\bm p, \bm p'} f_{\bm p'} \right),
\end{eqnarray}
where $e$ is the electric charge  and  $W_{\bm p', \bm p}$ is the probability for an electron to have the momenta
$\bm p$ before and $\bm p'$ after scattering. 
Lack of an inversion center for the surface charge carriers makes their elastic 
scattering asymmetric so that ${W_{\bm p \bm p'} \neq W_{-\bm p, -\bm p'} }$\cite{Belinicher-Strurman-UFN,GaN_LPGE}, 
and results in a \textit{dc} electric current.  

The total photocurrent density can be obtained from  the 
quadratic in the electric field stationary solution 
of Eq.~(\ref{BoltzmannPGE}) by applying the
standard equation ${\bm j = e\sum_{\bm p}  f_{\bm p} v_{\rm F} \bm p/p}$, where $v_{\rm F}$ is the
Fermi velocity. 
For elastic scattering by Coulomb impurities the photogalvanic 
current, e.g., in $x$-direction is given by\cite{Olbrich2014,Plank2017} 
\begin{eqnarray}
	\label{chi}
	j_x  = e v_{\rm F} {2\tau \over E_{\rm F}} \Xi \,\sigma(f) [|E_{x}|^2-|E_{y}|^2], 
\end{eqnarray}
where
$\Xi$ is the asymmetric scattering probability, $E_{\rm F}$ is the Fermi energy, $\tau$ is the scattering time, and $\sigma(f)$ is the high frequency (Drude) conductivity given by 
\begin{eqnarray}
\sigma(f) = \frac{e^2 E_{\rm F} \tau}{4\pi \hbar^2 [1+(2\pi f\tau)^2]}.
\label{Drude}
\end{eqnarray}
Measurements performed for frequencies ranging from fractions of THz up to 35~THz confirm
this frequency dependence, see Fig.~\ref{figure2}(a).
To cover this wide range of frequencies numerous sources of continuous wave 
($cw$) and pulsed infrared/terahertz laser radiation were applied including optically 
pumped molecular THz lasers\cite{LechnerAPL2009}$^-$\cite{DMSPRL09},
free electron lasers (FELBE) 
at the Helmhotz Zentrum Dresden-Rossendorf\cite{FEL1,FEL2}, as well as $Q$-switched and transversely excited 
atmospheric pressure (TEA) CO$_2$ lasers\cite{JETP1982,jiangPRB2010}. 
The lasers operated at single frequencies in the range from $f\approx 0.6$ to 35~THz 
(corresponding photon energies range from $\hbar \omega \approx  2.5$ 
to 145\,meV). 
In the majority of the experiments, the angle $\alpha$ was varied either by rotation of half-wave plates or of a grid wire placed behind a  quarter-wave Fresnel rhomb, which was set to  provide circularly polarized radiation\cite{PRB2007}. 

These spectra allow one to determine the scattering times of  Dirac fermions in the 
surface states despite the presents of thermally activate residual 
impurities in the material bulk.  The room temperature scattering  times in various materials scale from 0.04 to more than 0.25~ps and are summarized in Ref.~[11] 
for molecular beam epitaxy (MBE)  grown (Bi$_{1-x}$Sb$_x$)$_2$Te$_3$ based 3D TIs including a pure Bi$_2$Te$_3$ sample\cite{Plucinski2011,Kampmeier2015}, Bi$_2$Te$_3$/Sb$_2$Te$_3$  heterostructures with different thicknesses of the Sb$_2$Te$_3$  layer\cite{Eschbach2015,Lanius2016}, and (Bi$_{1-x}$Sb$_{x}$)$_2$Te$_3$  ternary systems\cite{Weyrich2016}. 
The samples, besides their composition, discriminate due to their Fermi level (E$_{\rm F}$) position measured by \textit{in-situ} ARPES or bulk carrier concentration. 
With knowledge of E$_{\rm F}$ the room  temperature mobilities are estimated with typical values, depending on the material,  from one to several thousand cm$^2$/Vs. 

Besides an access to the surface states carrier mobility, photogalvanic 
spectroscopy allows to map the orientation and distribution of the surface states 
domains.
Indeed the photogalvanic current is shown to be highly sensitive
to the relative orientation of the radiation electric field and
asymmetric scatters (wedges). 
The orientation of wedges can be extracted from the sign of the parameter $\chi$ as well as from the possible phase shift
in the polarization dependence caused by disorientation of the 
wedges and directions in which the LPGE current is measured, 
see Suppl. Mater. for Ref.~[8].
A detailed study of the electronic topography in 3D TIs has been 
reported in Ref.~[10].
Figure~\ref{figure2} (b)
shows the spatial variation of the LPGE magnitude obtained by
scanning the beam across the large area ($4\times 7$~mm$^2$) 
Bi$_2$Te$_3$ sample. The facts that the LPGE amplitude does not change 
its sign and in the measured polarization dependencies no phase shift has been detected 
clearly  indicate the identical
orientation of wedges at each point. 
This observation is supported by $X$-rays diffraction measurements yielding the average domain orientation. 
While the domains are shown to be equally oriented, 
the magnitude of the photocurrent varies 
from point to point and for certain areas approaches zero.
The latter can be attributed to  twisted domains\cite{Plank2016_2}. 

\vspace{0.5cm}
\textbf{4. Photocurrents excited at oblique incidence}
\vspace{0.1cm}

\begin{figure}[t]
	\includegraphics[width=\linewidth]{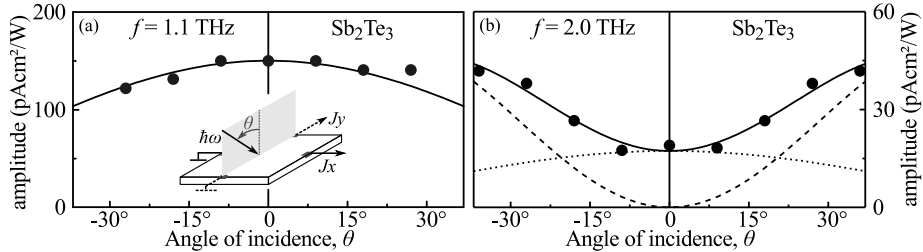}
	\caption{
		Dependence of the photocurrent amplitude on the angle of incidence $\theta$, measured in Sb$_2$Te$_3$ for excitation frequencies of (a) $f = 1.1$~THz and (b) $f = 2.0$~THz. 
		Fits after Eqs.~(\ref{chi}) and (\ref{oblique}). 
		Inset in panel (a) shows experimental setup. 
		After Ref.~[9]. 
	}
	\label{figure3}
\end{figure}
Now we briefly address the results on photocurrents generated at oblique incidence. 
Under this condition the LPGE photocurrent, described by Eq.~(\ref{normal}),
decreases as $\cos\theta$ due to reduction of the in-plane electric field components. 
Indeed, this behavior has been observed for some samples 
and radiation frequencies, see Fig.~\ref{figure3} (a).
This , however, is not always the case because in general oblique incidence may results in the generation of photogalvanic and/or photon drag effects caused by  microscopic mechanisms different from that considered above.  
New roots of the photocurrent emerge from the normal to the surface 
component of the radiation electric field, $E_z$, and an in-plane component of the photon momentum. 
The phenomenological theory 
developed for the surface states in Suppl. Mat. of  Ref.~[8] 
shows that oblique incidence can give rise to several types of LPGE and photon drag 
currents being either odd or even in the angle $\theta$. 
Consequently, an experimental evidence for a possible change of the mechanism responsible to the current formation
can be obtained by studying the angle of incidence dependence. 

Experiments on BiSbTe- based 3D TIs revealed that in the terahertz range 
oblique incidence causes the so-called trigonal photon drag effect, whereas
other roots of photocurrents play no essential role\cite{Plank2016}. 
The trigonal photon drag effect is even in the angle $\theta$ and results in an increase
of the photocurrent magnitude as compared to that of LPGE at normal incidence. This is shown
in Fig.~\ref{figure3} (b). Such a photocurrent behavior
is in agreement with phenomenological and microscopic theories developed in Ref.~[9].
The former one yields for the ($yz$) plane of incidence the photon drag current
\begin{eqnarray}
	 \label{oblique}
	j_x &=&      -  T_\parallel q_y E_y E_z  \\	
	j_y &=&     -  T_\parallel q_y E_x E_z,  \nonumber
\end{eqnarray}
with the in-plane photon momentum $q_y$. 
The developed microscopic model and theory show that the photon drag photocurrent is also caused by the \textit{dynamical} momentum alignment by time and space dependent radiation electric fields and implies the difference in the scattering probabilities for different half periods of the electromagnetic wave, for details see Ref.~[9]. 
The trigonal photon drag coefficient $T_\parallel$ obtained by solving the kinetic Boltzmann equation 
for oblique incidence and Coulomb impurity scattering is given by
\begin{eqnarray}
\label{T-PDE}
T_\parallel = \sigma(f) {e \beta \lambda^w p_{\rm F}^2 2\pi f\tau \over (2\pi f\tau)^2+9} \,. 
\end{eqnarray}
Here, $\beta$ is an anisotropic scattering constant, $p_{\rm F}$ is the Fermi momentum, and 
$\lambda^w$ is  a dimensionless degree of warping, which according to ARPES measurements can be of order of unity in our samples.
We emphasize that  both photocurrents, the trigonal photon drag effect at oblique incidence
and the LPGE at normal incidence,  stem from the same scattering events and, therefore, can be 
applied to study the high frequency conductivity in TI.

\vspace{0.5cm}
\textbf{5. Summary}
\vspace{0.1cm}

Recent studies of nonlinear electron transport phenomena in 3D and 2D TIs have already resulted in a great variety of fascinating robust effects. We still need to develop a full understanding of many of the effects using new experimental and theoretical concepts.
In particular, tuning the photocurrents by using external magnetic fields or strain will lead to new insights and mechanisms of the nonlinear transport in topologically protected states. 
Although photogalvanic effects were studied in a wide spectral range from terahertz to near infrared, 
the origin of photocurrents excited at high frequencies 
is not yet fully understood and await novel experimental and theoretical approaches. New roots of photogalvanic effects would provide studying of 3D TIs with a lateral superlattice as it has been demonstrated for structured graphene\cite{Koniakhin2014,ratchet_graphene} and two-dimensional gas in semiconductors\cite{Olbrich_PRB_11}.
Moving beyond topological insulators, we expect that similar effects can be studied in Weyl fermions\cite{Chan2016}$^-$\cite{ExpLee} 

and transition metal dichalcogenides 
(TMDC)\cite{Yuan2014}. 
Finally, we believe that the described effects will become  useful for material characterization as well as new non-linear devices based on topological insulators.

\vspace{0.5cm}
\textbf{6. Acknowledgment}
\vspace{0.1cm}

We thank L. E.  Golub for fruitful discussions. 
The support from the DFG priority program SPP 1666,
 the Elite Network of Bavaria (K-NW-2013-247), and the Volkswagen Stiftung Program
is gratefully acknowledged.


\begin{thebibliography}{99}
	
	\bibitem{Hasan} 
	M. Z. Hasan and C. L. Kane, 
	Rev. Modern Phys. {\bf 82}, 3045 (2010). 
	
	\bibitem{Qi}
	X. L. Qi and S. C. Zhang, 
	Rev. Mod. Phys. \textbf{83}, 1057 (2011).
	
	\bibitem{Ando}
	Y. Ando, 
	J. Phys. Soc. Japan {\bf  82}, 102001 (2013). 
	
	\bibitem{ARPES_1} 
	Y. Xia, D. Qian, D. Hsieh, L. Wray, A. Pal, H. Lin, A. Bansil, D. Grauer, Y. S. Hor, R. J. Cava, and M. Z. Hasan,
	Nature Physics \textbf{5}, 398 (2009).
	
	\bibitem{Bardarson2013} 
	J. H. Bardarson and J. E. Moore, 
	Rep. Prog. Phys. \textbf{76}, 056501 (2013).  
	
	\bibitem{GlazovGanichev_review} 
	M. M. Glazov and S. D. Ganichev,
	Phys. Reports \textbf{535},  101 (2014).
		
	\bibitem{Ivchenko2017}
	E. L. Ivchenko and  S. D. Ganichev, 
	Spin-dependent photogalvanic effects (A Review), 
	arXiv:1710.09223 (2017). 
		
	\bibitem{Olbrich2014} 
	P. Olbrich, L. E. Golub, T. Herrmann, S. N. Danilov, H. Plank, V. V. Bel'kov, G. Mussler, Ch. Weyrich, C. M. Schneider, J. Kampmeier, D. Gr\"{u}tzmacher, L. Plucinski, M. Eschbach, and S. D. Ganichev, 
	Phys. Rev. Lett. \textbf{113}, 096601 (2014). 
	
	\bibitem{Plank2016} 
	H. Plank, L. E. Golub, S. Bauer, V. V. Bel'kov, T. Herrmann, P. Olbrich, M. Eschbach, L. Plucinski, C. M. Schneider, J. Kampmeier, M. Lanius, G. Mussler, D. Gr\"{u}tzmacher, and S. D. Ganichev, 
	Phys. Rev. B \textbf{93}, 125434  (2016).
	
	\bibitem{Plank2016_2} 
	H. Plank, S. N. Danilov, V. V. Bel'kov, V. A. Shalygin, J. Kampmeier, M. Lanius, G. Mussler, D. Gr\"{u}tzmacher, and S. D. Ganichev, 
	J. Appl. Phys.  \textbf{120}, 165301 (2016). 
	
	\bibitem{Plank2017}
	H. Plank, J. Pernul, S. Gebert, S. N. Danilov, J. K\"{o}nig-Otto, S. Winnerl, M. Lanius, J. Kampmeier, G. Mussler, I. Aguilera, D. Gr\"{u}tzmacher, and S. D. Ganichev, 
	arXiv:1711.11457 (2017). 
		
	\bibitem{HosurBerry} 
	P. Hosur,  
	Phys. Rev. B {\bf 83}, 035309 (2011).
	
	\bibitem{Sheng2012} 
	Q. S. Wu, S. N. Zhang, Z. Fang, and Xi. Dai, 
	Physica E \textbf{44}, 895 (2012). 
	
	\bibitem{McIver2012} 
	J. W. McIver, D. Hsieh, H. Steinberg, P. Jarillo-Herrero, and N. Gedik, 
	Nature Nanotechn. \textbf{7}, 96 (2012).
	
	\bibitem{Duan2014} 
	J. Duan, N. Tang, X. He, Y. Yan, S. Zhang, X. Qin, X. Wang, X. Yang, F. Xu, Y. Chen, W. Ge, and B. Shen, 
	Scient. Rep. \textbf{4}, 4889 (2014).
	
	\bibitem{Junck2014} 
	A. Junck, G. Refael, and F. von Oppen, 
	Phys. Rev. B \textbf{90}, 245110 (2014).
		
	\bibitem{Hamh2016} 
	S. Y. Hamh, S.-H. Park, S.-K. Jerng, J. H. Jeon, S.-H. Chun, and J. S. Lee, 
	Phys. Rev. B \textbf{94}, 161405(R) (2016). 
	
	\bibitem{Okada2016} 
	K. N. Okada, N. Ogawa, R. Yoshimi, A. Tsukazaki, K. S. Takahashi, M. Kawasaki, and Y. Tokura, 
	Phys. Rev. B  \textbf{93}, 081403(R) (2016). 
	
	\bibitem{Pan2017}
	Y. Pan, Q.-Z. Wang, A. L. Yeats, T. Pillsbury, T. C. Flanagan, A. Richardella, H. Zhang, D. D. Awschalom, C.-X. Liu, and N. Samarth, 
	Nature Comm. \textbf{8}, 1037 (2017).
		
	\bibitem{Dantscher2017}  
	K.-M. Dantscher, D. A. Kozlov, M.-T. Scherr, S. Gebert, J. B{\"a}renf{\"a}nger, M. V. Durnev, S. A. Tarasenko, V. V. Bel'kov,   N. N. Mikhailov, S. A. Dvoretsky, Z. D. Kvon,  J. Ziegler, D. Weiss, and S. D. Ganichev,
	Phys. Rev. B Rapid Comm. \textbf{95}, 201103(R) (2017).	
		
	\bibitem{Eroms}  S.\,D.~Ganichev, D. Weiss, and J. Eroms,
	Annalen der Physik \textbf{529}, 1600406   (2017).
	
	\bibitem{Drexler2013} C. Drexler,
	S. A. Tarasenko, P. Olbrich, J. Karch, M. Hirmer, F. M\"{u}ller, M. Gmitra, J. Fabian, R. Yakimova, S. Lara-Avila, S. Kubatkin, M. Wang, R. Vajtai, P. M. Ajayan, J. Kono, and S. D. Ganichev, 
	Nature Nanotech. {\bf 8}, 104 (2013). 
	
	\bibitem{Dantscher2015} 
	K.-M. Dantscher, D. A. Kozlov, P. Olbrich, C. Zoth, P. Faltermeier, M. Lindner, G. V. Budkin, S. A. Tarasenko, V. V. Bel'kov, Z. D. Kvon, N. N. Mikhailov, S. A. Dvoretsky, D. Weiss, B. Jenichen, and S. D. Ganichev, 
	Phys. Rev. B \textbf{92}, 165314 (2015). 
		
	\bibitem{Ivchenkobook2} E. L. Ivchenko, 
	\textit{Optical Spectroscopy of Semiconductor Nanostructures} 
	(Alpha Science Int., Harrow, UK, 2005).
	
	\bibitem{book} S. D. Ganichev and  W. Prettl, 
	\textit{Intense Terahertz Excitation of Semiconductors}
	(Oxford Univ. Press, Oxford, 2006).
	
	\bibitem{Belinicher-Strurman-UFN} V. I. Belinicher and B. I. Sturman,
	Sov. Phys. Usp. \textbf{23}, 199 (1980).
	
	\bibitem{GaN_LPGE} W. Weber, L. E. Golub, S. N. Danilov, J. Karch, C. Reitmaier, B. Wittmann, V. V. Bel'kov, E. L. Ivchenko, Z. D. Kvon, N. Q. Vinh, A. F. G. van der Meer, B. Murdin, and S. D. Ganichev, 
	Phys. Rev. B \textbf{77}, 245304 (2008).
	
	\bibitem{LechnerAPL2009}  
	V. Lechner, L. E. Golub, P. Olbrich, S. Stachel, D. Schuh, W. Wegscheider, V. V. Bel'kov, and S. D. Ganichev,
	Appl. Phys. Lett. \textbf{94}, 242109 (2009).	
	
	\bibitem{Kvon2008}  Z. D. Kvon, S. N. Danilov, N. N. Mikhailov, S. A. Dvoretsky, and S. D. Ganichev,
	Physica E \textbf{40}, 1885 (2008). 
	
	\bibitem{DMSPRL09} S. D. Ganichev, S. A. Tarasenko, V. V. Bel'kov, P. Olbrich, W. Eder, D. R. Yakovlev, V. Kolkovsky, W. Zaleszczyk,
	G.~Karczewski, T. Wojtowicz, and D. Weiss,
	Phys. Rev. Lett. \textbf{102}, 156602 (2009). 	 
	
	\bibitem{FEL1} 
	P. Michel, F. Gabriel, E. Grosse, P. Evtushenko, T. Dekorsy, M. Krenz, M. Helm, U. Lehnert, W. Seidel, R. W\"{u}nsch, D. Wohlfarth, and A. Wolf, 
	Proc. FEL Conf., 8 (2004). 
	
	\bibitem{FEL2} 
	P. Michel, H. Buettig, F. Gabriel, M. Helm, U. Lehnert, Ch. Schneider, R. Schurig, W. Seidel, D. Stehr, J. Teichert, S. Winnerl, and R. W\"unsch, 
	The Rossendorf IR-FEL ELBE, Proc. FEL Conf., 488 (2006). 
	
	\bibitem{JETP1982} 
	S. D. Ganichev, S. A. Emel'yanov, and I. D. Yaroshetskii,
	JETP Lett. \textbf{35}, 368 (1982).
	
	\bibitem{jiangPRB2010}
	C. Jiang, V. A. Shalygin, V. Y. Panevin, S. N. Danilov,  M. M. Glazov, R. Yakimova, S. Lara-Avila, S. Kubatkin, and S. D. Ganichev, 
	Phys. Rev. B \textbf{84}, 125429 (2011).
	
	\bibitem{PRB2007}  S. D. Ganichev, S. N. Danilov, V. V. Bel'kov, S. Giglberger, S. A. Tarasenko, E. L. Ivchenko, D. Weiss, W. Jantsch, F. Sch\"{a}ffler, D. Gruber, and W. Prettl,
	Phys. Rev. B  \textbf{75}, 155317 (2007).
	
	\bibitem{Plucinski2011} 
	L. Plucinski, G. Mussler, J. Krumrain, A. Herdt, S. Suga, D. Gr\"utzmacher, and C. M. Schneider, 	
	Appl. Phys. Lett. \textbf{98}, 222503 (2011).
	
	\bibitem{Kampmeier2015} 
	J. Kampmeier, S. Borisova, L. Plucinski, M. Luysberg, G. Mussler, and D. Gr\"{u}tzmacher, 
	Cryst. Growth Des. \textbf{15}, 390 (2015).
	
	\bibitem{Eschbach2015} 
	M. Eschbach, E. Mlynczak,  J. Kellner, J. Kampmeier, M. Lanius, E. Neumann,	C. Weyrich,	M. Gehlmann, P. Gospodaric, S. D\"oring, G. Mussler,	N. Demarina, M. Luysberg, G. Bihlmayer, Th. Sch\"apers, L. Plucinski, S. Bl\"ugel,	M. Morgenstern, C. M. Schneider, and D. Gr\"utzmacher, 
	Nature Commun. \textbf{6}, 8816 (2015). 
	
	\bibitem{Lanius2016} 
	M. Lanius, J. Kampmeier, C. Weyrich, S. K\"olling, M. Schall, P. Sch\"uffelgen, E. Neumann, M. Luysberg, G. Mussler, P. M. Koenraad, Th. Sch\"apers, and D. Gr\"utzmacher, 
	Cryst. Growth Des. \textbf{16}, 2057 (2016). 
	
	\bibitem{Weyrich2016} 
	C. Weyrich, M. Dr\"ogeler, J. Kampmeier, M. Eschbach, G. Mussler, T. Merzenich, T. Stoica, I. E. Batov, J. Schubert, L. Plucinski, B. Beschoten, C. M. Schneider, C. Stampfer, D. Gr\"utzmacher, and Th. Sch\"apers, 
	J. Phys.: Condens. Matter  \textbf{28}, 495501 (2016). 
	
	\bibitem{Koniakhin2014} S. V. Koniakhin,
	Eur. Phys. J. B  \textbf{87}, 216 (2014).
	
	\bibitem{ratchet_graphene} P. Olbrich, J. Kamann, M. K\"{o}nig, J. Munzert, L. Tutsch, J. Eroms, D. Weiss, M.-H. Liu, L. E. Golub, E. L. Ivchenko, V. V. Popov, D. V. Fateev, K. V. Mashinsky, F. Fromm, Th. Seyller, and S. D. Ganichev, 
	Phys. Rev. B \textbf{93}, 075422 (2016). 
	
	\bibitem{Olbrich_PRB_11} P. Olbrich, J. Karch, E. L. Ivchenko, J. Kamann, B. M\"{a}rz, M. Fehrenbacher, D. Weiss, and S. D. Ganichev,
	{Phys. Rev. B} \textbf{83}, 165320 (2011). 	
	
	\bibitem{Chan2016} C.-K. Chan,  P. A. Lee,  K. S. Burch,  J. H. Han,  and Y. Ran, 
	Phys. Rev. Lett. \textbf{116}, 026805 (2016).
	
	\bibitem{Moore} F. de Juan, A. G. Grushin, T. Morimoto, and J. E. Moore, 
	Nature Commun. {\bf 8}, 15995 (2017).
	
	\bibitem{Patrick} C.-K. Chan, N. H. Lindner, G. Refael, and P. A. Lee, 
	Phys. Rev. B \textbf{95}, 041104 (2017).
	
	\bibitem{Koenig} E. J. K{\"o}nig, H.-Y. Xie, D. A. Pesin, and A. Levchenko,  
	Phys. Rev. B {\bf 96}, 075123 (2017). 
	
	\bibitem{spivak} L. E. Golub, E. L. Ivchenko, and B. Z. Spivak, 
	Pis'ma Zh. Eksp. Teor. Fiz. \textbf{105}, 744 (2007); JETP Lett. {\bf 105}, 782 (2017).
	
	\bibitem{ExpLee}  Q. Ma, S.-Y. Xu, C.-K. Chan, C.-L. Zhang, G. Chang, Y. Lin, W. Xie, T. Palacios, H. Lin, S. Jia, P. A. Lee, P. Jarillo-Herrero, and N. Gedik, 
	Nature Phys. \textbf{13}, 842 (2017).
	
	\bibitem{Yuan2014}  H. Yuan, X. Wang, B. Lian, H. Zhang, X. Fang, B. Shen, G. Xu, Y. Xu, S.-C. Zhang, H. Y. Hwang, and Y. Cui,
	Nature Nanotech. \textbf{9}, 851 (2014).
	
\end{thebibliography}
\end{document}